\documentclass[10pt]{article}
\usepackage[utf8]{inputenc}
\usepackage{mathptmx}
\usepackage[affil-it]{authblk} 
\usepackage[a4paper, margin=1in]{geometry}
\usepackage{sectsty}
\sectionfont{\fontsize{10}{10}\selectfont}
\subsectionfont{\fontsize{10}{10}\selectfont}
\usepackage{titlesec}
\titleformat*{\subsection}{\normalsize\itshape}
\usepackage{indentfirst}
\usepackage{array,multirow}
\usepackage{fancyhdr}
\usepackage{graphicx}
\usepackage{caption}

\title{\huge \textbf{Explicit behaviors affected by driver's trust in\\a driving automation system
}}

\author{
        \large \textbf{Hailong LIU~$^{*}$, Toshihiro HIRAOKA~$^{*}$, Seiya TANAKA~$^{\#}$}\\  
        $^{*}$~Institutes of Innovation for Future Society, Nagoya University\\
        $^{\#}$~School of Engineering, Nagoya University\\
        \ \\
        \normalsize
        Furo-cho, Chikusa-ku, Nagoya, Aichi, 464-8601, JAPAN\\
        Phone: (+81) 52-747-6977 \\
        E-mail: liu.hailong@mirai.nagoya-u.ac.jp \\
        Co-author’s E-mail: toshihiro.hiraoka@mirai.nagoya-u.ac.jp\\
        Co-author’s E-mail: tanaka.seiya@h.mbox.nagoya-u.ac.jp
        }
\date{}
\begin{document}
\maketitle
%\thispagestyle{fancy}
%\keywords{Human-machine system, Over-trust in automation, Driving automation system}

\section*{Abstract}
As various driving automation system~(DAS) are commonly used in the vehicle, the over-trust in the DAS may put the driver in the risk.
In order to prevent the over-trust while driving, the trust state of the driver should be recognized.
However, description variables of the trust state are not distinct.
This paper assumed that the outward expressions of a driver can represent the trust state of him/her-self.
The explicit behaviors when driving with DAS is seen as those outward expressions.
In the experiment, a driving simulator with a driver monitoring system was used for simulating a vehicle with the adaptive cruise control~(ACC) and observing the motion information of the driver. 
Results show that if the driver completely trusted in the ACC, then 
1) the participants were likely to put their feet far away from the pedals; 2) the operational intervention of the driver will delay in dangerous situations.
In the future, a machine learning model will be tried to predict the trust state by using the motion information of the driver.

\section{Introduction}
In recent years, the driving automation system~(DAS) has been installed in the vehicles along with the improvement of sensor accuracy, calculation processing speed, and the recognition performance by using artificial intelligence technology.
%A part of the ADASs are intended to improve the driving safety, and other part of the ADASs are intended to reduce the driving workload of the driver.
%For example, the adaptive cruise control~(ACC) operates the accelerator and brake of the vehicle in a driving situation.
%The lane keeping assist~(LKA) supports the steering wheel operation via the detected lane marker on the road by using a camera.
%Meanwhile, the automatic emergency braking system~(AEB) is used to break the vehicle automatically when a obstacle in the front.
The DAS controls the vehicle based on analyzing the information observed by sensors. 
However, the different DAS has different requirements for usage conditions, e.g., driving task and operational domain.
The {\it SAE international} defined the DAS into five levels~\cite{SAE_j3016_2016}.
The levels one to three of the DAS require the driver to be fully or partially involved in the driving tasks, e.g., parts of the vehicle control and emergency intervention.
The driver may easily trust in the DAS when the driver does not understand the functional boundaries of the DAS.
Especially when a situation is beyond what the DAS can respond, the issue of ``over-trust'' will put the driver in danger.
The reason of the over-trust could be considered in that the driver cannot appropriately understand the system design, purpose, mechanism, and ability of the DAS.
To prevent the driver's over-trust in the DAS is a big challenge and final goal of our study.

In our previous study~\cite{liu2018trust}, the over-trust in the levels one to three of the DAS is defined as that the DAS cannot respond to certain driving tasks while the driver trust it can.
Consequently, there are two judgment conditions of over-trust:
{\bf 1)~the driver is trusting in the DAS}, 
{\bf 2)~the DAS cannot respond to driving tasks}.
Based on this definition, a model of the over-trust mechanism was proposed in~\cite{liu2018trust}.
The driver's part in the proposed model is based on two theories --- the mental model~\cite{grosser2012mental}, and the risk homeostasis theory~\cite{wilde1982theory}.
The driver will construct a mental model of the DAS through repeated use.
This mental model explains a thought process of the driver about how the DAS are functioning in the real world.
The driver recognizes the situation of the DAS by feedback information from an HMI of the DAS and compares the predicted result by using the mental model with the actual situation of the DAS in order to adjust his/her trust state. 
After that, the driver combines the trust state for the DAS, the perceived hazards from the traffic environment and the vehicle, and the driving skill to evaluate the perceived risk that is subjective. 
According to the risk homeostasis theory~\cite{wilde1982theory},
the driver determines driving behavior by comparing the perceived risk with his/her acceptable risk level and controls the body to complete the driving maneuver.
It is called risk compensation behavior.
In the vehicle's part of the proposed model, the DAS uses a variety of sensors to observe the environmental information and the vehicle state information.
Above information is analyzed to generate control signals for supporting the driver to complete the driving tasks.
This paper considers the trust as a psychological activity, and the reliance as a behavioral manifestation of the trust.
The driver will take reliance behaviors when this driver trusts in DAS.
Moreover, if the DAS cannot respond the driving tasks, then the vehicle will out of control.
In this case, the driver is over-trusting in the DAS.

The final goal of this study is to prevent over-trust based on the above two conditions.
This paper is the first step towards our final goal--analyzing explicit behaviors when a driver is trusting the DAS.

%\begin{figure}[tb]
%\centering
%\includegraphics[width=1\textwidth]{FAST_ZERO_trust_model.pdf}
%\caption{The over-trust model proposed in the previous study.The task of this paper is shown in the red boxes.}
%\label{fig:model}
%\end{figure}

\section{Previous studies of the trust in automation}
Some early research on human-to-machine trust focused on the automated production lines that analyzed the relationship between the user's trust and strategy~\cite{Lee&Moray(1992),Muir&Moray(1996)}.
In the study field of human-robot interaction, the trust of human in robots is a hot research direction~\cite{desai2013impact,Geiskkovitch2019}.
With the automation of the vehicle, the problems of the driver’s trust in the automatic driving system were focused on.
Itoh investigated the changes of the driver's reliance when the driver trusted in an ACC going through multiple events~\cite{Itoh(2012)}.
He analyzed the velocity at braking, the time headway, and the time-to-collision to confirm that the ADAS was over-trusted by the drivers.
Abe et. al focused on the driver's trust in an automation vehicle when it was overtaking and passing~\cite{abe2017driver}.
The experimental result showed that the driver's trust in an automation vehicle may be affected by the velocity of ego vehicle, the lateral distance from the object, and steering manoeuvre start time when it was overtaking.
In the above studies, the short-term changes in the driver's trust could not be observed because participants were asked to evaluate their trust after each experiment.
In~\cite{desai2013impact}, they observed the trust in real-time but the trust states were taken in every 25 seconds.
In this paper, continuous data of the driver's trust state needs to be observed to investigate the effects of the driver's trust states on driver's behavior.
An evaluation method for continuously observing the driver's trust status was used in this study.

\section{Hypothesis}
In this paper, the effect on the driver’s motion under the driver's trust state was analyzed.
A great example of the level one DAS -- adaptive cruise control system~(ACC) is applied in experiments.
The ACC can operate the accelerator and brake of the vehicle in a driving situation.
Note that the ACC is designed to be difficult to react to stationary objects~\cite{ITOH2009}.
The driver's trust in the ACC may be easier established because ACC takes longer time to operate the vehicle than other ADASs (level one DASs).
We assume that there are different driving maneuvers for drivers with different trust state in the ACC.
For example, if the driver is completely trusting in ACC, then the driver’s right foot places next to the pedals, and the pedals are not operated.
If the driver is moderately trusting the ACC, then the driver’s right foot puts on the brake pedal, ready to intervene in the ACC.
If the driver is completely distrusting in ACC, then the driver presses on the brake pedal to cancel the ACC's control.
Therefore, there are two hypotheses of experiments as the following.
Based on the above assumption, the motions of the driver's right foot and leg can be used to represent the driver's trust state in ACC.
There are two hypotheses of experiments as the following.
\begin{description}
\item[Hypothesis 1:]~The driver will move the right foot away from pedals when the trust increases in ACC.
\item[Hypothesis 2:]~The high trust in DAS will delay operational intervention of the driver in dangerous situations.
\end{description}

\section{Overview of experiments}
\begin{figure}[tb]
\centering
\begin{minipage}[t]{0.48\textwidth}
\centering
\includegraphics[width=1\textwidth]{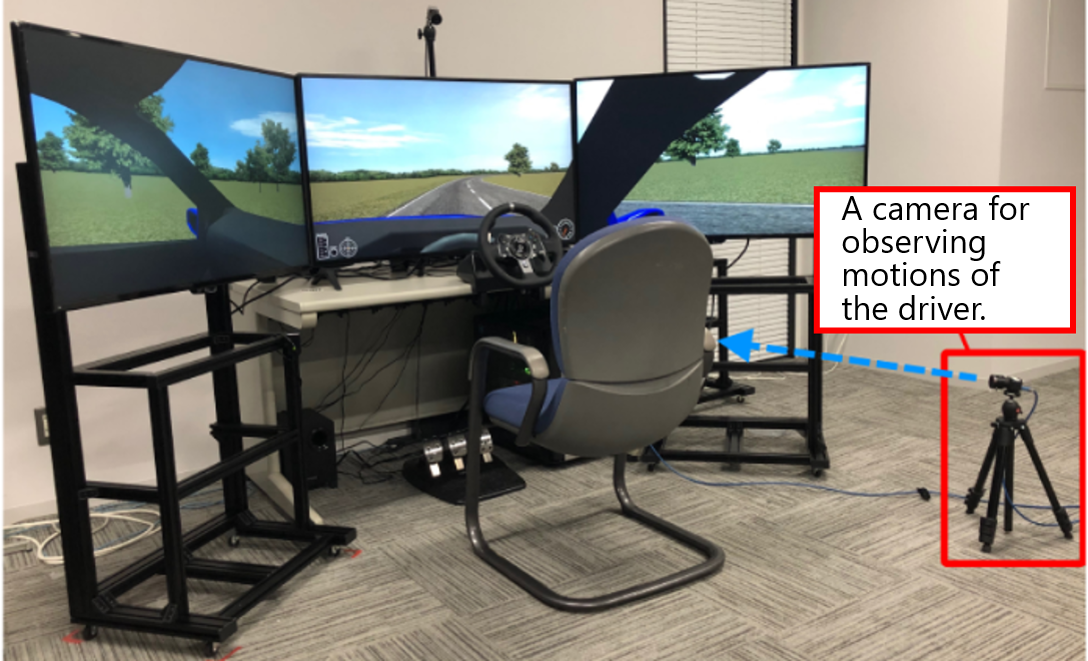}
\caption{Experimental installations: a driving simulator with a camera. The camera is used to observe the driver's motions.}
\label{f1}
\end{minipage}
\hspace{0.5mm}
\begin{minipage}[t]{0.48\textwidth}
\centering
\includegraphics[width=0.8\textwidth]{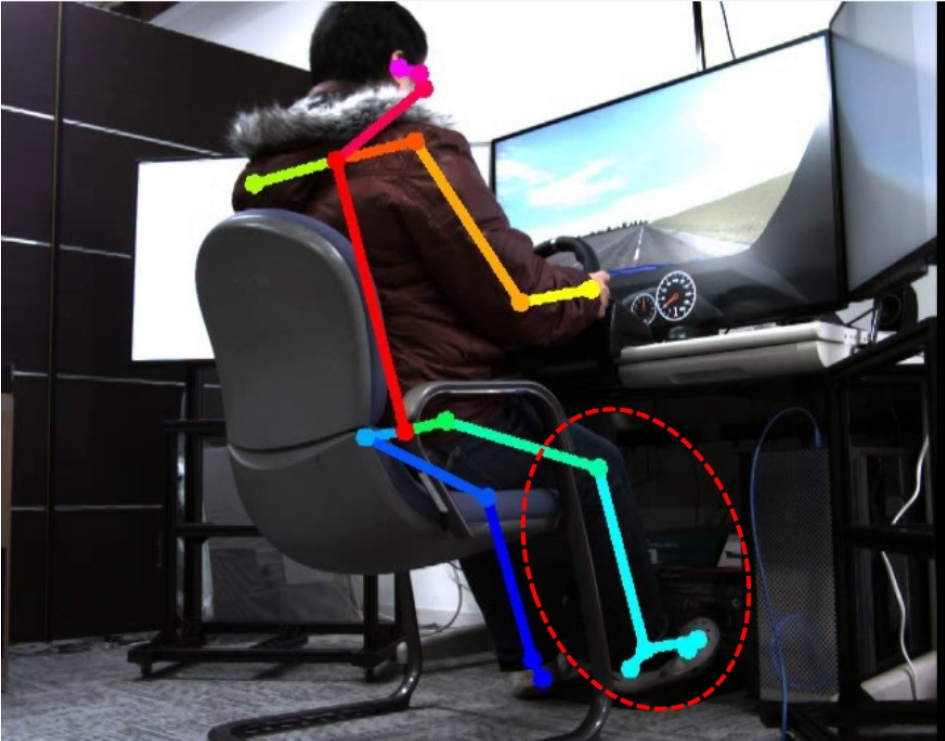}
\caption{Observing the motion information of the driver's right foot and leg by using {\it OpenPose}.}
\label{f2}
\end{minipage}
\end{figure}

The purpose of this experiment is to clarify the effect on the driver’s motion under the driver's trust states.
A virtual vehicle with ACC is implemented on a driving simulator~(DS).
Figure~\ref{f1} shows the driving simulator used in the experiment.
The DS bases on the {\it CarSim DS} ver.2018.1 with sensors option.
It has three 4K displays and a gaming steering-pedal controller.
In the DS, the ego vehicle runs at a constant speed $100~[km/h]$ by ACC when there is no preceding vehicle.
The following driving is carried out while keeping the target inter-vehicle distance when there is a preceding vehicle.
The target inter-vehicle distance is computed from the time headway ($2~[s]$) and the safety margin ($2~[m]$).
Note that the sensor of ACC is in the front of the vehicle at the center, and it's detection range was $100~[m]$ with $\pm1~[deg]$.
The {\it OpenPose}~\cite{cao2018openpose} is used to observe the motions of the driver's right foot and leg by using a camera~(Fig.~\ref{f1}).
The information of the joint points on the right leg and foot (in the red circle of Fig.~\ref{f2}) were used to analyze only.
The information for each joint point is a two-dimensional coordinate point that is the pixel location of the joint point at the image.

There were 13 participants with the ages from 19 to 50 (Avg:$24$ SD:$7.6$) attended the experiments.
All participants are Japanese and they have the driving license.
Participants are asked to evaluate their trust state in the ACC as a time series data through steering wheel paddles (Fig.~\ref{f3}) in real time while driving.
When the paddles are not touched, it indicates that the participants completely trusts in the ACC.
ACC satisfies the requirements for the driver to drive comfortably and safely, and the driver leaves the driving task to ACC with peace of mind.
If either side of the paddle is pressed, this indicates that the participant moderately trusts the ACC.
In this case, the participant cautiously grants the ACC control authority when the ACC is suspected that it does not perfectly perform the driving task for safety.
The participant completely distrusts in the ACC when he/she clearly believe that ACC cannot guarantee driving safety, both paddles will be pressed.
Noted that the above operation of the paddle does not press it once when the trust state changes, but needs to pressed it continuously while driving.
In order to reduce the impact of using steering wheel paddles on the participants' driving tasks, the driving scene is set on a straight highway with three lanes.
Multiple events of vehicle cutting-in occur during participants driving a vehicle with ACC.
Participants are required to do not change lanes but they allowed to intervene in ACC's control, i.e. acceleration, deceleration or turning off the ACC by the ACC button (Fig.~\ref{f3}), when they feel threatened.
Note that the ACC button can be used to turn the ACC on and off, participants can also turn off the ACC by pressing the brake pedal.
In addition, if the accelerator pedal is depressed, then the participant's operation takes priority when ACC is on.
If the accelerator pedal is released, the operation returns to the ACC automatically.
%%%%%%%%%%%%%%%%%%%%

Before the experiment, participants were asked to be familiar with the usage of DS and learn to evaluate their trust state in the ACC during driving in about 30 minutes.
There were three driving scenarios designed to experiment.
Each scenario took about 20 minutes.
In order to induce participants to trust the ACC, the first scenario included various events that could be well responded by the ACC, e.g., smooth cut-in four times, smooth lane change four times, and multiple smooth accelerations by the preceding vehicle.
In the second scenario, The behaviors of the preceding vehicle became more intense, which included four times cut-in and three times lane changes.
In these events, ACC can still safely control the vehicle.
It lets participants to understand the functional limits of ACC and to further build participant's trust in the ACC.
A dangerous event is set at the end of the third scenario.
The preceding vehicle -- a large trailer made a sharp turn to avoid a stopped vehicle in an emergency.
In this case, the ACC will difficult to respond to the stopped vehicle.
Therefore, the participant must be involved in the control of the vehicle in time.
After the experiment using the DS, a visit was conducted for each participant.

\section{Experiments result}
\begin{figure}[tb]
\begin{minipage}[t]{0.52\textwidth}
\centering
\includegraphics[width=0.6\textwidth]{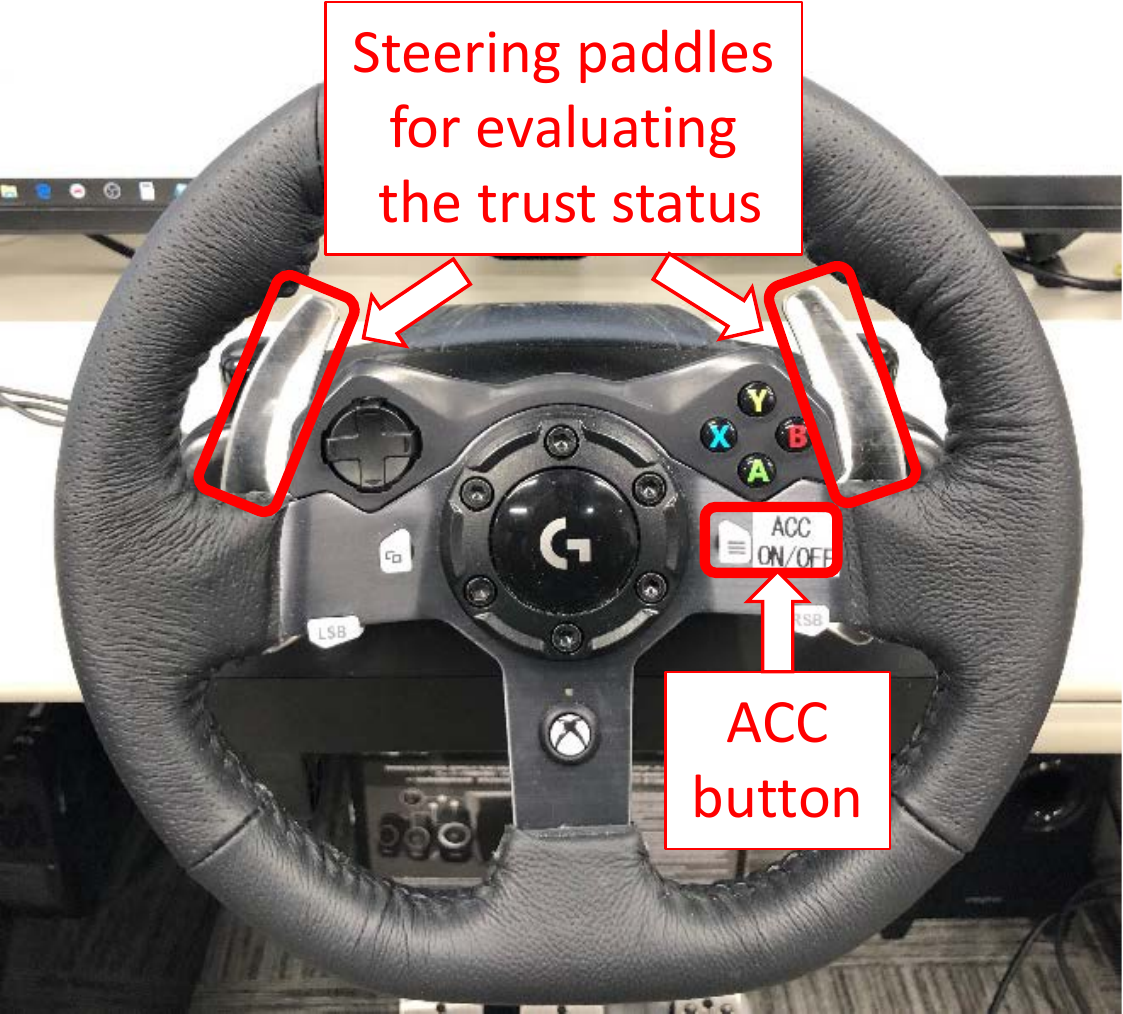}
\caption{The steering wheel with ACC button and paddles.}
\label{f3}
\end{minipage}
\hspace{0.5mm}
\begin{minipage}[t]{0.48\textwidth}
\centering
\includegraphics[width=1\textwidth]{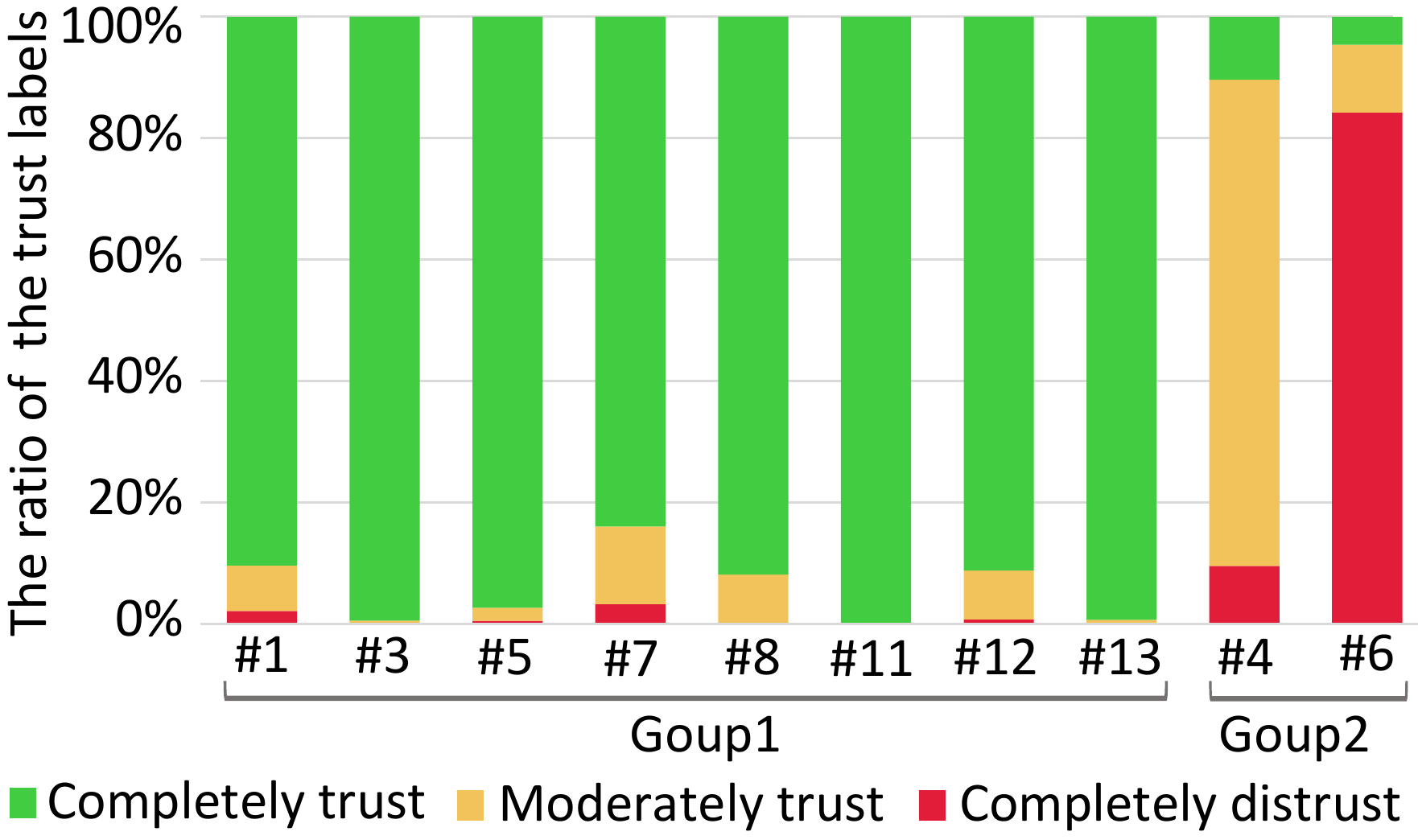}
\caption{Ratio of the trust labels for all scenarios.}
\label{Fig:trust_ratio}
\end{minipage}
%\hspace{0.7mm}
%\begin{minipage}[t]{0.196\textwidth}
%\centering
%\includegraphics[width=1\textwidth]{place.pdf}
%\caption{Definitions of the places.}
%\label{fig:place}
%\end{minipage}
%\hspace{0.5mm}
%\begin{minipage}[t]{0.386\textwidth}
%\centering
%\includegraphics[width=1\textwidth]{place_result.pdf}
%\caption{Rate of that drivers put right foot on each place.}
%\label{fig:place_result}
%\end{minipage}
\end{figure}

Data from 13 participants were observed.
However, data from three participants ($\#2$,$\#9$,$\#10$) was excluded for analysis because of equipment issues.
%\subsection{The ratio of the trust state}
Figure~\ref{Fig:trust_ratio} shows the ratio of the trust state for 10 participants in all scenarios.
The ratio of the trust state "completely trust" accounted for less than $10\%$ when the participants $\#4$ and $\#6$ drove the vehicle with the ACC.
Meanwhile, other participants accounted for more than $84\%$.
Through interviews, we found that participants $\#4$ and $\#6$ were very cautious in their daily driving.
In the experiment, they immediately reduce the trust in the ACC when some tiny risk occurred.
A cautious attitude would make them difficult to regain complete trust in ACC.
Therefore, participants $\#4$ and $\#6$ are classified as the group~2, and the other participants ($\#1$, $\#3$, $\#5$, $\#7$, $\#8$, $\#11$ and $\#13$) as the group~1.

%\subsection{The ratio of the right foot position}
%Nine position of right foot were defined as Fig~\ref{fig:place}.
%The positions of the participant's right foot were manually labeled in time series by checking the video of the experiment.
%For safety using the ACC, the foot should be placed on A or B. 
%This allows for faster intervention in ACC control when the danger comes.
%However, the foot will become stiff and uncomfortable when the foot is putted on the pedal for a long time.
%Before the experiment, we predicted that participants had a high probability of placing their right foot on the %AF, BF or AS when using ACC long time considering comfort and safety.
%However, Fig.~\ref{fig:place} shows that participants' right foot position were significantly different when using ACC.
%For example, participants $\#1$ and $\#13$ always placed their right foot on either pedal.
%Meanwhile, participant $\#8$ placed his right foot on O2 that was far from the pedal.
%Through interviews, we found that each participant's confidence in their ability to intervene in ACC was very different.
%It may lead to they putted their right foot in the different positions.
%This result has caused us to ponder: where should the driver put the foot when using ACC? 
%Should the layout of pedals be designed for driving safety and comfortably?

\subsection{Verification for the hypothesis~1}

We considered that participants strategies were similar in the same trust state although each participant had their own strategies of putting the right foot on different places when using the ACC.
Recalling the hypothesis~1 is that the driver moves the right foot away from pedals when the trust increases in ACC.
In order to verify the hypothesis~1, a principal component analysis~(PCA) was used to reduce the dimensions of motion data observed by {\it OpenOpse} for visualization.

\begin{figure}[tb]
\centering
\includegraphics[width=1\textwidth]{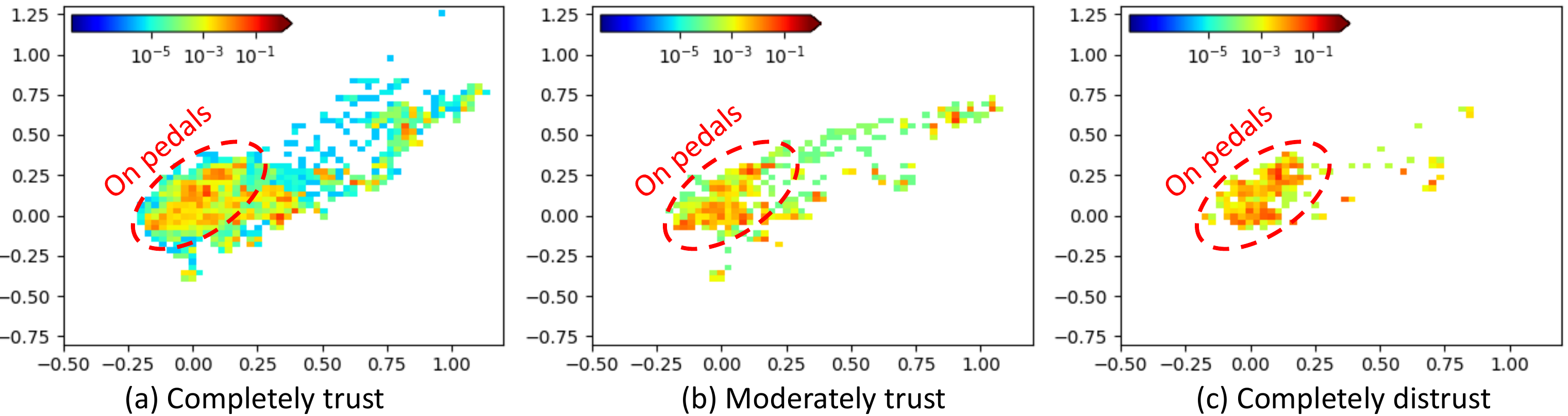}
\caption{The two dimensional features of group 1 by PCA based on the principal component of \#12's data.}
\label{fig:PCA_G1}
\vspace{2mm}
\centering
\includegraphics[width=1\textwidth]{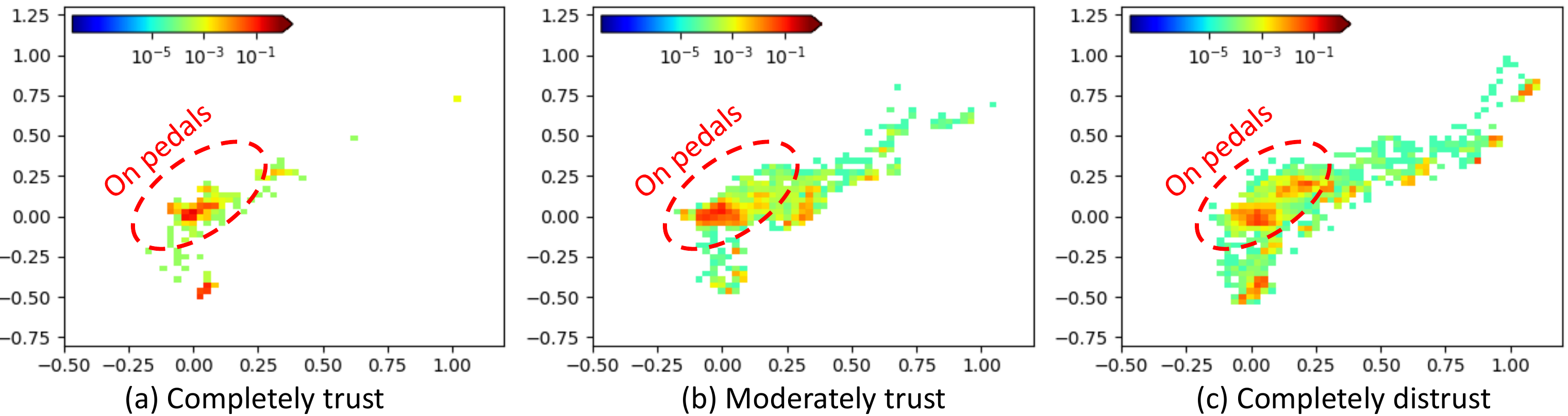}
\caption{The two dimensional features of group 2 by PCA based on the principal component of \#12's data.}
\label{fig:PCA_G2}
\end{figure}

The transformation of participant $\#12$ was used to project the motion data into a two-dimensional space for all the participants because the $\#12$’s two-dimensional features could represent his the position of the right foot more clearly than others'.
The position of the foot in three different trust state is discussed separately.
The result of the group~1 and 2 are shown in Fig.~\ref{fig:PCA_G1} and \ref{fig:PCA_G2}.
The color of the dots represents the probability density of foot positions. 
If a position is placed on the foot many times, the color will be close to red, otherwise, the color will be close to blue.
For the group~1 (Fig.~\ref{fig:PCA_G1}), regardless of their trust or distrust in the ACC, they mostly put their feet on the pedals.
There was a tendency that their feet were not only on the pedals but also far away from the pedals when they completely trusted in the ACC.
While, when they completely distrusted in the ACC, the positions of their feet were almost on the pedals (Fig.~\ref{fig:PCA_G1} (c)).
It showed that participants were likely to put their feet far away from the pedals when they completely trusted in the ACC although this result only partially matched the hypothesis~1.

The results of the group~2 are shown in Fig.~\ref{fig:PCA_G2}, which were completely opposite to the results of the group~1.
The difference between cautious participants and others can be seen although the number of participants in the second group was small.
%It is difficult to explain this result.
We could only assume that the cautious attitude made the two participants balance their mental preparation and physical preparation for intervening the ACC.
For example, when they completely trusted in the ACC, although the mental preparation may be low, the physical preparation may be high because the right foot placed on the pedals.
Meanwhile, their right foot was not placed on the pedal in some cases when they didn't trust in the ACC.
In these cases, their physical preparation may be low, but the mental preparation may be high, that is, mentally ready to intervene.
This balancing process of mental and physical preparation may be related to risk compensation behavior.
This assumption needs further verification in the future.

\subsection{Verification for the hypothesis 2}
\begin{figure}[tb]
\centering
\begin{minipage}[t]{0.32\textwidth}
\centering
\includegraphics[width=1\textwidth]{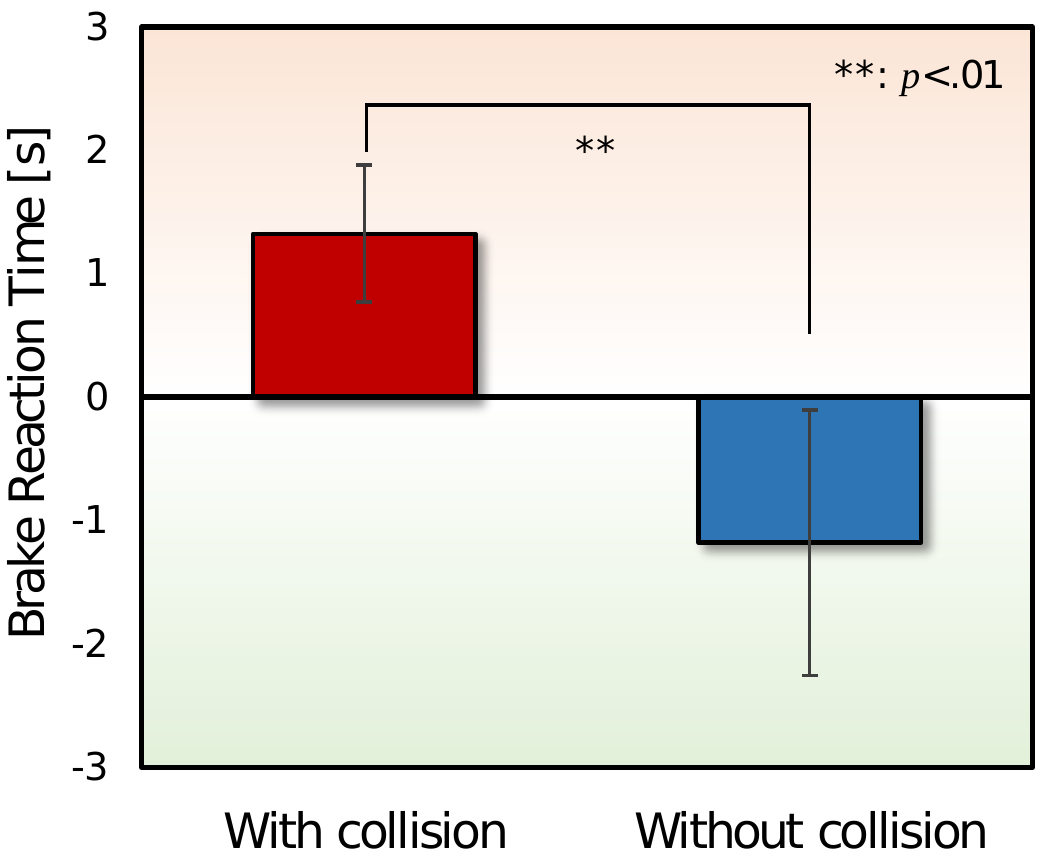}
\caption{The difference of the brake reaction time between the collision group and the non-collision group in the dangerous event.}
\label{fig:collision_vsBrake}
\end{minipage}
\hspace{0.5mm}
\begin{minipage}[t]{0.32\textwidth}
\centering
\includegraphics[width=1\textwidth]{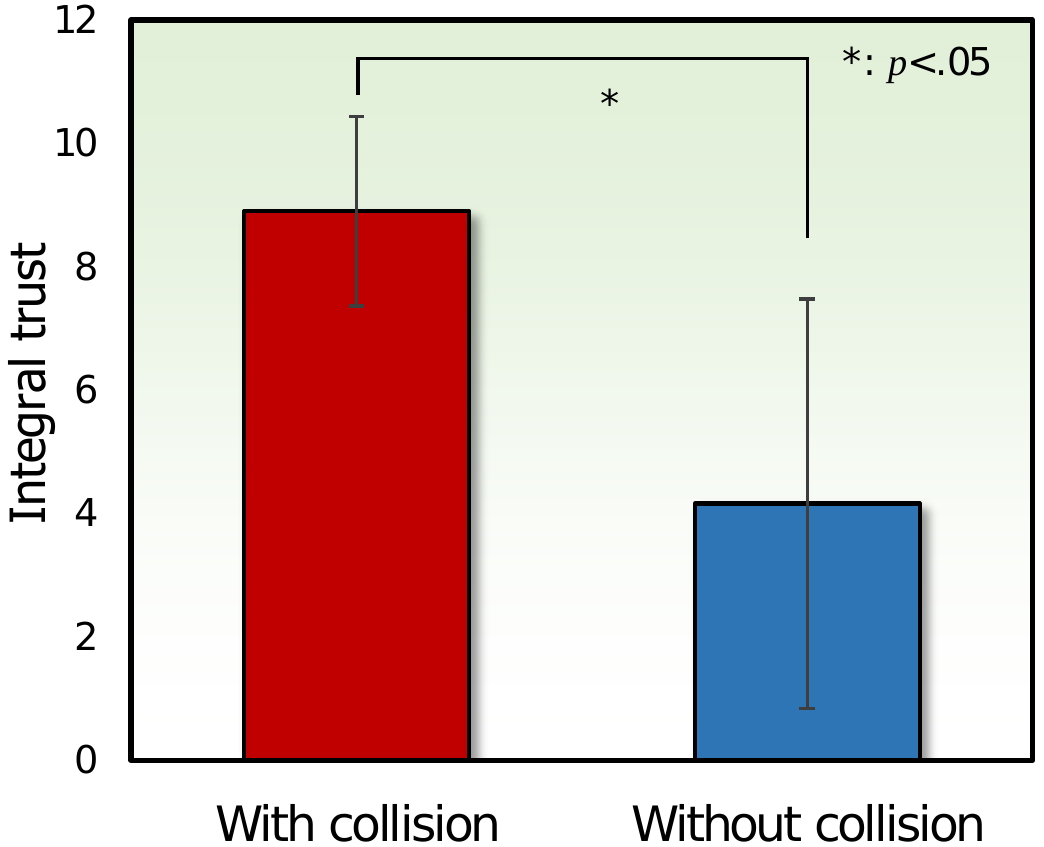}
\caption{The difference of the integral trust between the collision group and the non-collision group in the dangerous event.}
\label{fig:collision_vsTrustE}
\end{minipage}
\hspace{0.5mm}
\begin{minipage}[t]{0.32\textwidth}
\centering
\includegraphics[width=1\textwidth]{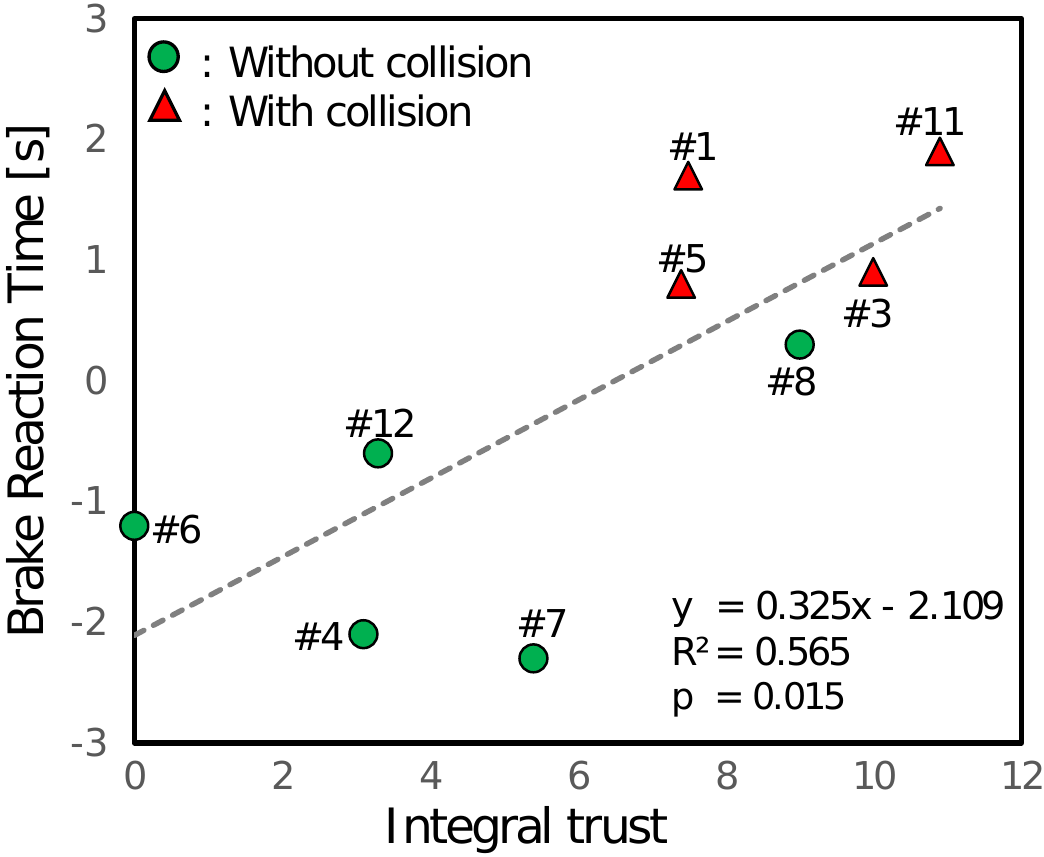}
\caption{The relationship between the integral trust and the brake reaction time in the dangerous event.}
\label{Fig:TrustVSBrake}
\end{minipage}
\end{figure}

The over-trust in DAS was defined as when the driver trusts in the DAS, but the DAS does not guarantee driving safety.
According to this definition, we investigated the driver's trust state in the ACC and the reaction time of intervening control when ACC was a failure.
Recalling the hypothesis~2: higher trust in DAS delays operational intervention in dangerous situations.
A dangerous event that cannot be responded by ACC was set at the end of the third scenario.
The start time point of the dangerous event is defined when the trailer begins to decelerate rapidly that was 3.2 seconds before the trailer's lane change.
Meanwhile, the critical point of brake reaction time is defined as when the trailer went out of the ACC's detection range.
At this time, the stopped vehicle before the trailer would be revealed.
In this dangerous event, half of the participants collided who were $\#1$, $\#3$, $\#5$, $\#11$ and $\#13$.
The participants $\#4$, $\#6$, $\#7$, $\#8$ and $\#12$ stepped on the brakes pedal in time so that there were no collisions.
Not that the participant $\#13$ did not step on the brake pedal until the collision.
Thus, the data for $\#13$ were excluded in the following analysis of the brake reaction time.

The Fig.~\ref{fig:collision_vsBrake} shows the brake reaction time of the collision group and non-collision group in the dangerous event.
The vertical axis represents time, and the zero-point represents the critical point of brake reaction time.
A plus value indicates that the brake reaction time is later than the critical point, and the minus value indicates that it is earlier than the critical point.
The brake reaction time of the collision group (without $\#13$) was later than the non-collision group.
A T-test was used to verifies the significantly different between those two groups.
The corresponding p-value was 0.01, indicating a significant difference in the incidence between collision group (without $\#13$) and non-collision group.

In order to count the participants’ trust state in the ACC, the trust state in the interval of the dangerous event were integrated, where ``completely trust'' was 2, ``moderately trust'' was 1, and ``completely distrust'' was 0 at each time step. 
The difference of the integral trust between the collision group (with $\#13$) and the non-collision group is shown in Fig.~\ref{fig:collision_vsTrustE}.
The integral trust of the collision group (with $\#13$) was significantly higher than it of the non-collision group with the p-values 0.05 by a T-test.
This result also directly indicates that the trust states of the participants were related to the collision occurrence.

The relationship between brake reaction time and the integral trust was investigated for verifying the hypothesis~2.
The data of the collision group (without $\#13$) and the non-collision group were used in this verification.
The results are shown in Fig.~\ref{Fig:TrustVSBrake} that there was a high correlation between them, where the correlation coefficient was 0.75, and the p-value from a test for no correlation was 0.015.
A linear regression function was used to map the integral trust to the brake reaction time, and the coefficient of determination~($R^2$) of this linear regression function was 0.565.
The Fig.~\ref{Fig:TrustVSBrake} clearly shows that the brake reaction time and integral trust of the non-collision group are lower than the collision group except participant $\#8$.
Although the participant $\#8$ had a higher integral trust than $\#1$ and $\#5$, and the brake reaction time was later than $\#4$, $\#6$, $\#7$ and $\#12$, she used a very heavy brake to stop the ego vehicle within a distance of less than half a meter from the stop vehicle.

The above results clearly show that hypothesis~2 is established.
The high trust in DAS will delay operational intervention of the driver in dangerous situations.
Furthermore, according to the experiment results and the definition of over-trust, if a driver higher trusts in ACC when the ACC cannot respond to the driving tasks, then the probability of an accident will higher.

%In order to verify this assumption, we designed an experiment.

%Meanwhile, the driver's motions will be observed by the camera with the {\it OpenPose}.
%A principal component analysis~(PCA) is asked to extract the low-dimensional latent features from the observed high-dimensional motion data.
%It helps to visualize the high-dimensional data.

%\subsection{Result}
%By comparing results of PCA and video recording, it shows that if the driver highly trusted in the ACC, his/her foot was farther from the pedals.
%Besides, the results show that more highly the drivers trusted in ACC , more later they reacted to intervene in dangerous event.
%Through the above machine learning model, it is possible in real-time to predict the driver's trust state in the ACC.

\section{Conclusion}
This paper discussed the mechanism of driver's over-trust in the DAS.
To prevent the over-trust while driving, finding the explicit behaviors of a driver which can represent the trust state of him/her-self is a very important task.
For this task, a DS with a driver monitoring system was used for simulating a vehicle with the ACC and observing the motion information of the driver.
Results show that if the driver completely trusted in the ACC, then 
1) the participants were likely to put their feet far away from the pedals; 2) the operational intervention of the driver will delay in dangerous situations.
In the future, a machine learning model will be tried to predict the trust state through the motion data of the driver.

\bibliographystyle{ieeetr}			
\bibliography{citation}
\end{document}